\documentclass[mathleft]{an}
\usepackage{graphicx}
\usepackage{times}
\begin{document}

\Pagespan{789}{}
\Yearpublication{2006}%
\Yearsubmission{2006}%
\Month{}%
\Volume{}%
\Issue{}%

\title{About universes with scale-related total masses and their abolition of presently outstanding cosmological problems}

\author{H.J. Fahr\inst{1}\fnmsep\thanks{Corresponding author: hfahr@astro.uni-bonn.de}
\and M. Heyl\inst{2}}

\titlerunning{About universes with scale-related total masses}
\authorrunning{H.\,J. Fahr \& M. Heyl}
\institute{ Argelander Institut fuer Astronomie, University of Bonn, Auf dem
H\"{u}gel 71, 53121 Bonn, Germany
\and Deutsches Zentrum f\"{u}r Luft und Raumfahrt (DLR),
K\"{o}nigswintererstra{\ss}e 522 - 524, 53227 Bonn, Germany}

\received{8 Mar 2006}
\accepted{\textbf{date to be included}}
\publonline{\textbf{date to be included}}


\abstract{The most recently celebrated cosmological implications of the cosmic
microwave background studies with WMAP (2006), though fascinating by
themselves, do, however, create some extremely hard conceptual challenges for
the presentday cosmology. These recent extremely refined WMAP observations
seem to reflect a universe which was extremely homogeneous at the
recombination age and thus is obviously causally closed at the time of the
cosmic recombination era. From the very tiny fluctuations apparent at this
early epoch the presently observable nonlinear cosmic density structures can,
however, only have grown up, if in addition to a mysteriously high percentage
of dark matter an even higher percentage of dark energy is admitted as drivers
of the cosmic evolution. The required dark energy density, on the other hand,
is nevertheless 120 orders of magnitude smaller then the theoretically
calculated value. These are outstanding problems of present day cosmology onto
which we are looking here under new auspices. We shall investigate in the
following, up to what degree a universe simply abolishes all these outstanding
problems in case it reveals itself as universe of a constant total energy. As
we shall show basic questions like: How could the gigantic mass of the
universe of about $10^{80}$ proton masses at all become created? - Why is the
presently recognized and obviously indispensable cosmic vacuum energy density
so terribly much smaller than is expected from quantumtheoretical
considerations, but nevertheless terribly important for the cosmic evolution?
-Why is the universe within its world horizon a causally closed system? - ,
can perhaps simply be answered, when the assumption is made, that the
universe has a constant total energy with the consequence that the total mass density of the universe (matter and vacuum) scales with $R_\mathrm{u}^\mathrm{-2}$. Such a scaling of matter and vacuum energy abolishes the horizon problem, and the cosmic vacuum energy density can easily be reconciled with its theoretical expectation values. In this model the mass of the universe increases linearily with the world extension $R_\mathrm{u}$ and can grow up from a Planck mass as a vacuum fluctuation.}

\maketitle

\section{Introduction}
This paper aims at the investigation of possible cosmological
consequences of the hypothesis that the total mass density - or equivalently
the energy density, of the universe (baryonic and dark matter, vacuum energy,
photons) behaves according to a $R_\mathrm{u}^{-2}$-law where $R_\mathrm{u}$ denotes the cosmic scale or, if extended to the light horizon, the extension
of the universe. In this respect $R_\mathrm{u}$ can be called the distance
between two arbitrary space points comoving with the Robert- son-Walker-like
homologous cosmological expansion of the universe, validating the relation
$\dot{R_\mathrm{u}}/R_\mathrm{u}=\dot{S}/S$, if $S$ for instance is the distance
between any two freely comoving galaxies (dots on top of quantities mean
derivatives with respect to cosmic time. This relation was already suggested
by considerations carried out by Kolb (1989), Overduin \& Fahr (2001, 2003) or
Fahr \& Heyl (2006). A detailed justification for such a scaling behaviour
will be provided in section 9 at the end of this paper where we will show that
such a density scaling indeed might help to describe a reasonably realistic
scenario which correctly delivers the basic parameters of our expanding universe.

In the following we shall work in the frame of General Relativity and shall
base ourselves on the Friedmann equations. The only special prerequisites for
our investigation are the assumption of a total mass density scaling with
$R_\mathrm{u}^\mathrm{-2}$ and a topologically flat universe with a curvature parameter $k=0$,
which is also strongly suggested nowadays by the recent WMAP data (WMAP 2006).

A short suggestive substantiation for our assumption of a $R_\mathrm{u}^\mathrm{-2}$-scaling can
be derived from the 1. Friedmann equation

\setlength{\mathindent}{0pt}
\begin{equation}
H^\mathrm{2}  = \left( {\frac{{\dot R_\mathrm{u}}}{R_\mathrm{u}}} \right)^\mathrm{2}  = \frac{{8\pi G\rho _\mathrm{tot}}}{3}  - \frac{kc^\mathrm{2}}{{R_\mathrm{u}^\mathrm{2}}}, \label{1}%
\end{equation}
with $H$ the Hubble parameter, $R_\mathrm{u}$ and $\dot{R_\mathrm{u}}$ the extension of the universe and its time derivative, respectively, $G$ the constant of gravitation, $c$ the velocity of light, and $\rho_\mathrm{tot}$ the total mass density of all gravitating constituents of the universe (e.g. baryonic and dark matter, equivalent mass of the vacuum energy). For a flat universe, $k=0$, the above equation can be written as

\setlength{\mathindent}{0pt}
\begin{equation}
\dot {R_\mathrm{u}}^\mathrm{2}  = \frac{{8\pi G\rho _\mathrm{tot}}}{3} R_\mathrm{u}^\mathrm{2} = \Phi_\mathrm{eff}, \label{2}%
\end{equation}
where the expression $8\pi G\rho _\mathrm{tot}R_\mathrm{u}^\mathrm{2}/3$ can be interpreted as an effective cosmic action potential $\Phi_\mathrm{eff}$ of gravitation which is the driving quantity for the dynamics of the universe. Respecting the cosmological principle means, that the potential $\Phi_\mathrm{eff}$ is present at each space point of the expanding universe. We now take as our basic assumption that the gravitational energy $\mathrm{d}m\Phi_\mathrm{eff}$, related to each mass element $\mathrm{d}m$ of our homogeneously matter-filled universe, just equals its rest mass energy, i.e.

\setlength{\mathindent}{0pt}
\begin{equation}
\Phi_\mathrm{eff} \mathrm{d}m = \frac{{8\pi G\rho _\mathrm{tot}}}{3} R_\mathrm{u}^\mathrm{2} \mathrm{d}m = c^\mathrm{2} \mathrm{d}m. \label{3}%
\end{equation}
Then, this evidently results in a total density $\rho_\mathrm{tot}$ given by

\setlength{\mathindent}{0pt}
\begin{equation}
\rho_\mathrm{tot}={\frac{{3c^\mathrm{2}}}{{8\pi GR_\mathrm{u}^\mathrm{2}}}} \propto {\frac {1}{R_\mathrm{u}^\mathrm{2}}}, \label{4}%
\end{equation}
which expresses the assumed $R_\mathrm{u}^\mathrm{-2}$-scaling as the basis of our further investigations. A deeper view to this relation is offered in section 9 of this paper.

\section{The radius of the universe}
First we consider the total mass of the universe $M_\mathrm{tot}$, i.e. the total cosmic mass contained within the spacelike 3-sphere of a radius equal to the light horizon $R_\mathrm{u}=c/H$, where $H$ denotes the Hubble constant, and introduce the mass-associated Schwarzschild radius $R_\mathrm{S}$ given by

\setlength{\mathindent}{0pt}
\begin{equation}
R_\mathrm{S}={\frac{2GM_\mathrm{tot}}{{c^\mathrm{2}}}}. \label{5}%
\end{equation}
The total mass of the universe contained in the spacelike 3-sphere of
radius $R_\mathrm{u}$ is obtained by adding up rest mass energies according
to the following expression

\setlength{\mathindent}{0pt}
\begin{equation}
M_\mathrm{tot}(t)c^\mathrm{2} =\int^{V_\mathrm{u}^\mathrm{{3}}}\rho%
_\mathrm{0}(t)c^\mathrm{2}\sqrt{-g_{3}}\mathrm{d}^\mathrm{3}V. \label{6}%
\end{equation}
Here $\rho_\mathrm{0}(t)$ denotes the density of homogeneously
distributed cosmic mass which is variable with cosmic time $t$, and
$\mathrm{d}^{3}V_{0}=\sqrt{-g_{3}}\mathrm{d}^\mathrm{3}V$ is the local differential proper
volume of space given through the determinant of the 3-d part of the
Robertson-Walker metric tensor in the form

\setlength{\mathindent}{0pt}
\begin{equation}
\sqrt{-g_\mathrm{3}} = \sqrt {-g_\mathrm{11}g_\mathrm{22}g_\mathrm{33}} = \sqrt {\frac{{R_u^\mathrm{6} }}
{{(1 - \frac{{kr^\mathrm{2} }}{4})^\mathrm{6} }}}  = \frac{{R_\mathrm{u}^\mathrm{3} }}{{(1 - \frac{{kr^\mathrm{2} }}{4})^\mathrm{3}}} \label{7}%
\end{equation}
with $k$ being the curvature parameter and $r$ the normalized radial coordinate of a polar coordinate system. Thus one
obtains the total mass of the universe in the form

\setlength{\mathindent}{0pt}
\begin{equation}
M_\mathrm{tot}(t)c^\mathrm{2} = 4\pi R_\mathrm{u}^\mathrm{3}(t)\rho%
_\mathrm{0}(t)c^\mathrm{2}\int_\mathrm{0}^\mathrm{1}\frac{r^\mathrm{2}\mathrm{d}r}{(1-\frac{kr^\mathrm{2}}{4})^\mathrm{3}}. \label{8}%
\end{equation}
For a universe with vanishing curvature, i.e. for $k=0$, this
simplifies to

\setlength{\mathindent}{0pt}
\begin{equation}
M_\mathrm{tot}(t) = \frac{4\pi}{3}R_\mathrm{u}^\mathrm{3}(t)\rho_\mathrm{0}(t). \label{9}%
\end{equation}
The above relation just expresses that, judged by a central observer
within a homogeneous and cosmologically expanding universe, the total mass
$M_\mathrm{tot}$ turns out to be a constant, unless for some yet unexplained
reason the rest mass density of the universe does not fall off according to $R_\mathrm{u}^\mathrm{-3}$.

The above expression permits to solve Eq. (\ref{5}) with
respect to the density $\rho_\mathrm{tot}$ which then yields

\setlength{\mathindent}{0pt}
\begin{equation}
\rho_\mathrm{tot}={\frac{{3c^\mathrm{2}}}{{8\pi G}}}{\frac{{R_\mathrm{S}}}{{R_\mathrm{u}^\mathrm{3}}}}. \label{10}%
\end{equation}
This result can only be reconciled with our assumption that the
density of the universe scales with $R_\mathrm{u}^\mathrm{-2}$, if the extent or the radius of
the universe just is the above defined Schwarzschild radius, i.e. if $R_\mathrm{u}=R_\mathrm{S}$,
then leading to the following relation

\setlength{\mathindent}{0pt}
\begin{equation}
\rho_\mathrm{tot} = {\frac{{3c^\mathrm{2} } }{{8\pi GR_\mathrm{u}^\mathrm{2} }}} = {\frac{{3c^\mathrm{2} } }{{8\pi
GR_\mathrm{S}^\mathrm{2} }}}, \label{11}%
\end{equation}
which is, by the way, identical to the result of Eq. (\ref{4}). From this above relation (\ref{11}) one can draw the conclusion that the
universe investigated has a world radius $R_\mathrm{u}$ which is identical with its
Schwarzschild radius $R_\mathrm{S}$ defined by its total mass $M_\mathrm{tot}$.

\section{The expansion velocity of the universe}
Now we investigate the characteristic expansion velocity for an
universe with $\rho_\mathrm{tot}$ given by Eqs. (\ref{4}) or (\ref{11}). With $k=0$ this velocity can be easily derived from the first Friedmann equation (\ref{1})

\setlength{\mathindent}{0pt}
\begin{equation}
\left( {\frac{{\dot R_\mathrm{u}}}{R_\mathrm{u}}} \right) ^\mathrm{2}  = \frac{8\pi G}{3}\rho _\mathrm{tot} = \frac{8\pi G}{3} {\frac{{3c^\mathrm{2} } }{{8\pi GR_\mathrm{u}^\mathrm{2} }}} =\frac {c^\mathrm{2}}{R_\mathrm{u}^\mathrm{2}}, \label{12}
\end{equation}
which then simply yields

\setlength{\mathindent}{0pt}
\begin{equation}
\dot{R_\mathrm{u}}^\mathrm{2} = c^\mathrm{2} \Rightarrow R_\mathrm{u} = c \Rightarrow H = \frac {c}{R_\mathrm{u}}. \label{13}
\end{equation}
As seen from the above relation, the expansion velocity of a universe with $\rho_\mathrm{tot} \sim R_\mathrm{u}^\mathrm{-2}$ equals the velocity of light $c$, results which so far remind to those obtained for a coasting universe with a constant expansion velocity presented by Kolb (1989).

\section{Density and mass of the universe}

With the above result for the Hubble parameter $H=c/R_\mathrm{u}$ the total density can be written as

\setlength{\mathindent}{0pt}
\begin{equation}
\rho_\mathrm{tot}= {\frac{{3c^\mathrm{2}}}{{8\pi GR_\mathrm{u}^\mathrm{2}}}} = {\frac{{3H^\mathrm{2}}}{{8\pi G}}}=\rho_\mathrm{crit}, \label{14}%
\end{equation}
meaning that the density $\rho_\mathrm{tot}$ permanently equals the critical density $\rho_\mathrm{crit}$, as should be the case for an uncurved universe with $k=0$. Replacing further on the density by the mass using the relation
$\rho_\mathrm{tot}=M_\mathrm{tot}/({\frac{4\pi}{3}}R_\mathrm{u}^\mathrm{3})$, one then is lead to the result

\setlength{\mathindent}{0pt}
\begin{equation}
M_\mathrm{tot}={\frac{{c^\mathrm{2}}}{{2G}}}R_\mathrm{u}, \label{15}%
\end{equation}
which states the perhaps astonishing fact that the total mass of the universe we have considered scales with its cosmic radius $R_\mathrm{u}$. The latter fact by the way has been emphazised as true for completely different reasonings, eg. by Einstein (1920), Dirac (1937), Whitrow (1946), Hoyle (1990, 1992), Fahr (2006), Fahr \& Heyl (2006) and Fahr \& Zoennchen (2006).

The above derived result finally challenges to put the question of how large the mass of the
universe under these auspices might have been at the very early cosmic time,
when the radius of the universe amounted to one Planck scale
$R_\mathrm{Pl}=\sqrt{hG/c^\mathrm{3}}$ of a Planck mass
$m_\mathrm{Pl}=\sqrt{\hbar c/G}$. Then, with
Eq. (\ref{15}) one obtains the following answer to this question,

\setlength{\mathindent}{0pt}
\begin{equation}
M_\mathrm{tot}(R_\mathrm{Pl})= \frac {1}{2} m_\mathrm{Pl}. \label{16}%
\end{equation}
This means that a universe with $\rho \sim R_\mathrm{u}^\mathrm{-2}$, even of the size
of the present universe with its gigantic mass, can possibly have evolved at
its expansion from half a Planck mass which according to quantummechanical results
can appear at any time on the Planck scale as a virtual mass fluctuation, i.e.
as a quantum fluctuation (see also section 7). In view of the required scaling of the cosmic mass with the scale $R_\mathrm{u}$ the present world mass amounts to

\setlength{\mathindent}{0pt}
\begin{equation}
M_\mathrm{tot}={\frac{{c^\mathrm{2}}}{{2G}}}R_\mathrm{u}={\frac{{c^\mathrm{3}}}{{2GH}}}\approx10^\mathrm{53}\mathrm{kg}\approx10^{80}m_\mathrm{prot}, \label{17}%
\end{equation}
where we have used $R_\mathrm{u}=c/H$ and a Hubble parameter of $H=\mathrm{72km/s/Mpc}$ for the present universe, or
synonymous, a present radius of the universe $R_\mathrm{u}$ amounting to 4167 $\mathrm{Mpc}$
($m_\mathrm{prot}$ is the mass of a proton). The above values are consistent with
standard estimations, e.g. what concerns the visible universe to consist of
about $10^\mathrm{11}$ galaxies with $10^\mathrm{11}$ solar-type stars each.

We conclude that the investigated universe can easily evolve from a quantummechanical fluctuation
which is allowed on the Planck scale $R_\mathrm{Pl}=\sqrt{hG/c^\mathrm{3}}$, followed by an ongoing materialization of vacuum energy with the required rate (section 7). Its matter density thereby permanently represents the critical density $\rho_\mathrm{crit}$.

\section{The horizon problem}
A big cosmological problem consists in the so-called "horizon problem"
that arises from the question, whether or not a photon released at the very
beginning of the universe, "t=0", i.e. the Big Bang, can meanwhile have
reached any potential observer or cosmic space point. This question leads one
to the introduction of the term "light horizon", reflecting the maximum
distance from any observer's space point up to which this observer can be
informed about physical events in the universe, since photons are assumed to
communicate such past events to the observer till today. On the other hand,
regions of the universe which are located behind this light horizon are thus
causally decoupled from the observer.

The distance of the light horizon for a universe with curvature $k=0$ can be
calculated as (see e.g. Stephani 1988)

\setlength{\mathindent}{0pt}
\begin{equation}
r_\mathrm{hor}=c\int\limits_\mathrm{0}^{t_\mathrm{univ}}{{\frac{{\mathrm{d}\tilde{t}}}{R_\mathrm{u}({\tilde{t})}}}}, \label{18}%
\end{equation}
with $R_\mathrm{u}(\tilde{t})$ denoting the time dependent extension of the universe.
Cosmic expansions with $R_\mathrm{u}\sim t_\mathrm{univ}^{\alpha}$ and $\alpha<1$ yield a finite
$r_\mathrm{hor}$ and therefore yield a causally unclosed universe - the sword of
Damocles of the todays cosmology, since to the contrast for instance the
cosmic microwave background with its origin about 150 kiloyears after the
Big-Bang is clearly indicating causal closure. In contrast to standard
cosmological solutions with $\alpha<1$, the expansion of the universe considered here 
simply follows the law $R_\mathrm{u}=ct_\mathrm{univ}$, i.e. requiring $\alpha=1!$ , and hence the
light horizon $r_\mathrm{hor}$ in this case is given by

\setlength{\mathindent}{0pt}
\begin{equation}
r_\mathrm{hor}=c\int\limits_\mathrm{0}^{t_\mathrm{univ}}{{\frac{{\mathrm{d}\tilde{t}}}{R_\mathrm{u}({\tilde{t})}}=}%
}\int\limits_\mathrm{0}^{t_\mathrm{univ}}{{\frac{{\mathrm{d}\tilde{t}}}{{\tilde{t}}}=\infty}}. \label{19}%
\end{equation}
Therefore we obtain the interesting result that an economical universe is
causally closed over the whole period of its expansion - or with other words:
All space points of an economical universe could have undergone a physical
interaction with each other, since at any time following the start of the
expansion the causal light horizon was infinite.

Hence we can draw the conclusion, that an universe with a density scaling according to $R_\mathrm{u}^{-2}$ does not
encounter a horizon problem and that no inflation is required to explain
the highly pronounced isotropy of the 2.735 $\mathrm{K}$ cosmic microwave background
(CMB) observed by WMAP.

\section{The problem with the vacuum energy}
According to the present observations the mass equivalent of the cosmic vacuum energy is
expected to contribute by about $70\%$ to the total mass of the universe (Perlmutter et al. 1999; WMAP 2006).
However, the vacuum energy as it is calculated by quantumfield theoreticians
confronts us with the problem that it amounts to values higher by a
factor $\approx10^\mathrm{120}$ compared to the value it should have to be conciliant 
with the nowadays observational results of the universe. This problem vanishes in the case that the mass or energy density of the universe, including the vaccum energy, follows the assumed $R_\mathrm{u}^\mathrm{-2}$ law.  By the way, a scaling of the vacuum energy density with $R_\mathrm{u}^\mathrm{-2}$ has already been speculated on at many other places in the literature (for a review see Overduin \& Cooperstock (1998)).

We start with a look on the vacuum energy which is generally interpreted
as the overall sum of the energy zero-point oscillations with proper
frequencies $\omega_\mathrm{j}$ of the vacuum. This sum can be expressed by the
following equation

\setlength{\mathindent}{0pt}
\begin{equation}
E_\mathrm{vac} = {\frac{1 }{2}}\sum\limits_\mathrm{j} {\hbar w_\mathrm{j}}. \label{20}%
\end{equation}
The corresponding density of the vacuum energy is then calculated to yield
(Weinberg 1989)

\setlength{\mathindent}{0pt}
\begin{equation}
\rho_\mathrm{vac}c^\mathrm{2}={\frac{c{\hbar k_\mathrm{\max}^\mathrm{4}}}{{16\pi^\mathrm{2}}}}, \label{21}%
\end{equation}
with $\hbar$ the Planck constant and $k_\mathrm{max}$ the so-called "cut-off" wave
number of the oscillations. Such a "cut-off" is needed to prevent
divergencies when calculating the energy density in the frame of QED. Since
the wave number $k_\mathrm{max}$ can be expressed by the Planck mass $m_\mathrm{Pl}%
=\sqrt{\hbar c/G}$ as $k_\mathrm{max}= m_\mathrm{Pl}c/\hbar$, the
vacuum energy density is obtained in the form

\setlength{\mathindent}{0pt}
\begin{equation}
\rho_\mathrm{vac}c^\mathrm{2}=\frac {c^\mathrm{7}}{{16\pi^\mathrm{2}\hbar G^\mathrm{2}}} \approx 10^\mathrm{120} \rho_\mathrm{crit} c^\mathrm{2}. \label{22}%
\end{equation}
In spite of the above equation, we now want to ask for an upper limit of cosmic vacuum energy as it would be given by an universe with a scaling density according to $R_\mathrm{u}^\mathrm{-2}$. Since we know that $\rho_\mathrm{vac}$ should not exceed the scaling critical density $\rho_\mathrm{crit}$,

\setlength{\mathindent}{0pt}
\begin{equation}
\rho_\mathrm{vac}c^\mathrm{2}\approx\rho_\mathrm{crit}c^\mathrm{2}={\frac{{3H^\mathrm{2} } }{{8\pi G }}}c^\mathrm{2} =
{\frac{{3c^\mathrm{4}}}{{8\pi GR_\mathrm{u}^\mathrm{2}}}}, \label{23}%
\end{equation}
we are able to provide a formula for such a limit. We calculate the vacuum energy
density with Eq. (\ref{23}) at a phase of the evolution when $R_\mathrm{u}=R_\mathrm{Pl}=\sqrt{\hbar G/c^\mathrm{3}}$
and we obtain instead of Eq. (\ref{22})

\setlength{\mathindent}{0pt}
\begin{equation}
\rho_\mathrm{vac}c^\mathrm{2}={\frac{{3c^\mathrm{7} } }{{8\pi\hbar G^\mathrm{2}}}}, \label{24}%
\end{equation}
which provides, using the Planck mass $m_\mathrm{Pl}=\sqrt{\hbar c/G}$ and the Planck volume
$V_\mathrm{Pl}=$ ${\frac{4}{3}}\pi R_\mathrm{Pl}^\mathrm{3}$, the following upper limit
for the density of the vacuum energy

\setlength{\mathindent}{0pt}
\begin{equation}
\rho_\mathrm{vac}c^\mathrm{2}={\frac{1 }{2}}{\frac{{m_\mathrm{Pl} c^\mathrm{2} } }{{\left(  {{\frac{4 }%
{3}}\pi R_\mathrm{Pl}^\mathrm{3} } \right)}}}. \label{25}%
\end{equation}
From this it can be concluded that the maximum density of the vacuum energy at
the very beginning of the universe obviously seems to be nothing else but the
energy density of half the Planck rest mass, or in terms of energy:
$E_\mathrm{vac}=\rho_\mathrm{vac}c^\mathrm{2}{\frac{4}{3}}\pi R_\mathrm{Pl}^\mathrm{3}={\frac{1}{2}}%
m_\mathrm{Pl}c^\mathrm{2}$. A comparison with Eq. (\ref{20}) then yields the surprising result

\setlength{\mathindent}{0pt}
\begin{equation}
{\frac{1 }{2}}m_\mathrm{Pl} c^\mathrm{2} = {\frac{1 }{2}}\sum\limits_\mathrm{j} {\hbar\omega_\mathrm{j}}. \label{26}%
\end{equation}
The above finding hence suggests the conclusion - that had already been
expressed as a presumption very often in the literature: The vacuum energy and
the equivalent energy of the Planck rest mass are in fact identical. In the
frame of an universe with a scaling density it now turns out, that the vacuum energy (=
Planck mass) can in fact be the source of all energy and matter in the universe.

Then, if one further on believes in the correctness of the presently favoured
fraction $\rho_\mathrm{vac}/\rho_\mathrm{mat}\simeq0.7/0.3$ of the vacuum energy and matter
density, respectively, one can also conclude that for some reason about $70\%$
of the total energy permanently remains in the vacuum during the expansion of
the universe - representing itself as vacuum energy - while about $30\%$ manifest itself as matter. This ratio must be
constant during the whole evolution of the universe because both, vacuum
energy and matter density, follow the assumed $R_\mathrm{u}^\mathrm{-2}$ scaling.

In this context it is interesting to have a short look at the well-known Casimir pressure that results from the interaction of a pair of neutral, parallel and electrically conducting planes due to the disturbance of the vacuum of the
electromagnetic field. The Casimir pressure is a pure quantum effect and is given by

\setlength{\mathindent}{0pt}
\begin{equation}
p_\mathrm{cas} = \frac{{\pi^\mathrm{2} \hbar c}}
{{240d^\mathrm{4} }}, \label{27}
\end{equation}
with $d$ being the distance between the planes. The expected Casimir pressure in the Planck region for the special case of two parallel conducting planes, i.e. $d=R_\mathrm{Pl}=\sqrt{\hbar G/c^\mathrm{3}}$, would then yield

\setlength{\mathindent}{0pt}
\begin{equation}
p_\mathrm{cas} = \frac{{\pi^\mathrm{2} c^\mathrm{7}}}
{{240\hbar G^\mathrm{2} }}. \label{28}
\end{equation}
On the other hand, the pressure associated with the vaccum energy density given by Eq. (\ref{24}) can be written as

\setlength{\mathindent}{0pt}
\begin{equation}
p_\mathrm{vac}=-\frac{1}{3}\rho_\mathrm{vac}c^\mathrm{2}=-{\frac{{c^\mathrm{7} } }{{8\pi\hbar G^\mathrm{2}}}}. \label{29}%
\end{equation}
We now calculate the ratio of both, the Casmir and vacuum energy pressure, and finally get

\setlength{\mathindent}{0pt}
\begin{equation}
\left|\frac{p_\mathrm{cas}}{p_\mathrm{vac}}\right| =  \frac{{8\pi^\mathrm{3} }}{{240}} \approx 1.033\ \approx 1. \label{30}
\end{equation}
We strictly point out, that this surprising result must not be overinterpreted, since the value of the Casimir pressure as well as its algebraic sign not only strongly depends on the shape of the interacting surfaces (e.g. plane-plane, plane-sphere, sphere-sphere, etc.) but also on the temperature. Furthermore, from cosmological point of view, also the topology of the space plays an important role (Bordag et. al. 2001). The above result, however, simply demonstrates that the Casimir pressure can be expected in the same order of magnitude as the associated pressure of the vacuum energy that we found to be the source of all matter in the universe. Thus, we conclude that the Casimir energy and the vacuum energy investigated in this paper must be somehow related with each other and that the expansion of the universe indeed might be based on the Casimir effect.

\section{Interpretation of the mass increase}
A universe with linearly increasing mass can be easily explained in the frame of quantum mechanics. A look at the uncertainty principle ${\hbar/2} \approx \Delta E\Delta t$ shows the possibility of the virtual apperarance of half a Planck mass within a time interval $\Delta t = t_\mathrm{Pl}$, i.e. Planck time

\setlength{\mathindent}{0pt}
\begin{equation}
{\frac {\hbar} {2}} \approx \Delta E\Delta t = \Delta mc^\mathrm{2} t_\mathrm{Pl}  = \Delta mc^\mathrm{2} \sqrt {{\frac {G\hbar} {c^\mathrm{5}}}}, \label{31}
\end{equation}
and thus

\setlength{\mathindent}{0pt}
\begin{equation}
\Delta m = {\frac {1} {2}}\sqrt {{\frac {\hbar c} {G}}}  = {\frac {1} {2}}m_\mathrm{Pl}. \label{32}%
\end{equation}
The virtual Planck mass may stay in the real world of the universe if its rest mass energy is compensated to zero. This could be guaranteed by the negative gravitational binding energy which each additional mass is subject to in the expanding universe and may lead - as shown in the next chapter - to a vanishing total energy of the whole universe. Thus, the mass increase of the universe could have its reason in virtual Planck masses which become real, perhaps similar to the way discussed by Hawking (1975) where real masses are ejected from the Schwarzschild boundary of black holes. This realized and materialized quanta contribute over the lifetime $t_\mathrm{univ}$ of the universe to the total mass $M_{tot}$ with a "production rate" of half a Planck mass each time interval $t_\mathrm{Pl}$ which leads to

\setlength{\mathindent}{0pt}
\begin{equation}
M_\mathrm{tot}  = {{\frac {1} {2}}{\frac {m_\mathrm{Pl}} {t_\mathrm{Pl}}}}t_\mathrm{univ} = {\frac {1} {2}}{\frac {c^\mathrm{3}} {G}}{\frac {R_\mathrm{u}}{c}} = {\frac {c^\mathrm{2}} {2G}}R_\mathrm{u}. \label{33}%
\end{equation}
Here we have used the result that the extension $R_\mathrm{u}$ of the investigated universe is simply given by $R_\mathrm{u}=ct_\mathrm{univ}$. It is amazing to recognize an identical scaling law for the mass as given in Eq. (\ref{15}). Obviously, quantum mechanical considerations on one hand and independently retrieved results for an universe with $\rho_\mathrm{tot} \sim R_\mathrm{u}^\mathrm{-2}$ on the other hand are in accordance. Furthermore, if we calculate the time derivative of the Schwarzschild radius in Eq. (\ref{5}) we retrieve

\setlength{\mathindent}{0pt}
\begin{equation}
\dot R_\mathrm{S}  = \frac{{\mathrm{d}R_\mathrm{S} }}{{\mathrm{d}t}} = \frac{{2G}}{{c^\mathrm{2} }}\frac{{\mathrm{d}M_\mathrm{tot} }}{{\mathrm{d}t}} = \frac{{2G}}{{c^\mathrm{2}}}\frac{{\left( {\frac{1}{2}m_\mathrm{Pl} } \right)}}{{t_\mathrm{Pl}}} = c. \label{34}
\end{equation}
Since the Schwarzschild radius, as shown earlier, represents the extension on the investigated universe the above result again indicates an expansion velocity $c$.

\section{Mass generation from vacuum energy}
We have shown above that an $R_\mathrm{u}^\mathrm{-2}-$decay of cosmic vacuum energy density can conciliate the completely disjunctive claims for very high vacuum energy densities from field theoreticians on the one hand with those
claims for very low vacuum energy densities originating from more recent results of observational cosmology on the other hand. Requiring a mass generation rate of half a Planck mass per Planck time for the expanding universe can nicely explain the present universe as coming from nothing, i.e. from pure vacuum conditions, and explains the adopted behaviour of both cosmic mass density and cosmic vacuum energy density as scaling with $R_\mathrm{u}^\mathrm{-2}$.

The question perhaps remains, why cosmic spacetime expansion should trigger the vacuum energy decay, and how the conversion of vacuum fluctuations into real massive particles may occur. This profound question has not been answered in this paper and up to now in fact has not been answered at all in a satisfactory manner in the scientific literature. Nevertheless there exist strong reasons to believe that cosmic vacuum energy, if it has an accelerating action on cosmic spacetime, has to decay. A constant vacuum energy doing an action on space by accelerating its expansion, without itself being acted upon, does not seem to be a concept conciliant with basic physical principles. In this respect all conservative Lambda-cosmologies taking $\rho_{vac}$ as a constant in our view are nonconvincing solutions. This has
meanwhile been realized by many authors (e.g. Fischer 1993; Massa 1994; Wetterich 1995; Overduin \& Cooperstock 1998; Fahr 2004, 2006) where variable-Lambda cosmologies have been discussed.

The connection of vacuum energy and matter creation has already been discussed in papers by Hoyle \& Narlikar (1966a, 1966b) and later by Hoyle (1990, 1992) and Hoyle, Burbidge \& Narlikar (1993). The requirement that general relativistic field equations should be conformally invariant with respect to any scale recalibrations leads these authors to the introduction of a general relativistic action potential which describes mass generation at
geodetic motions of particles. Connected with this mass generation a so-called $C$-field (creation-field) can be introduced which turns out to be connected with geodetic mass motions themselves. It can then be shown (Hoyle et al. 1993) that this $C$-field when introduced into the general relativistic field equations leads to terms equivalent to those resulting from vacuum energy. A similar connection between vacuum energy
density and mass density was also found by Massa (1994) who shows that Lambda should be proportional to mass density. Concluding these considerations one might say that up to now there is a lack of a rigorous formulation for the
transition of vacuum fluctuations into real masses, nevertheless there is at least the idea first discussed by Hawking (1975) that treating the quantummechanics of particles and antiparticles in the neighborhood of blackholes reveals the appearance of real particles, i.e. the decay of a pure vacuum into a mass-loaded vacuum around these blackholes. The so-called Hawking radiation in this respect is nothing else but a materialisation of vacuum energy in strong gravitational fields. Perhaps in this respect the expanding universe also represents a form of a time-dependent gravitational field which induces matter creation through the embedded quantummechanical wavefunctions of particles.

\section{Possible reason for the density scaling - The economical universe}
Several new ideas have been recently proposed which could help to
overcome the presently manifest cosmological problems. Besides the theory of
cosmic inflation which goes back to Guth \& Steinhardt (1984), Liddle \& Lyth (2000),
Guth (1999), or Linde (1984), some authors (e.g. Albrecht \&
Magueijo 1998; Fritsch 2002; Barrow 2003) have postulated a speed of
light variable with cosmic scale or time which would solve the flatness
problem and the horizon problem for cosmology. However, the problem connected with the
cosmologically required dark energy remains completely unsolved so far.

Already at earlier occasions the concept of an economical
universe with vanishing total energy which can be created from nothing but a quantum
fluctuation was introduced and discussed in first steps by Brout, Englert \& Gunzig (1978), Vilenkin (1982) or Tryon
(1973), and it may even go back to Jordan (1947). More recently, however, we
have revisited their ideas and have rederived a new form of this economical
universe (Overduin \& Fahr 2001, 2003; Fahr 2004). In
these latter publications it was formulated as a crucial requirement for an
economical universe, that its total energy $L=L(R_\mathrm{u})$ be minimal und equal to a
constant $L_\mathrm{0}$, or in other words requiring that the change of $L=L(R_\mathrm{u})$ with
the world radius $R_\mathrm{u}$, or with the cosmic evolution time $t$, vanishes at all
times of the cosmic evolution.

This requirement is expressed by the following relation

\setlength{\mathindent}{0pt}
\begin{equation}
L=E+U=L_\mathrm{0}, \label{35}%
\end{equation}
with

\setlength{\mathindent}{0pt}
\begin{equation}
E={\frac{{4\pi}}{3}}R_\mathrm{u}^\mathrm{3}(\rho_\mathrm{tot} c^\mathrm{2}+3p_\mathrm{tot}). \label{36}%
\end{equation}
In the above equation $\rho_\mathrm{tot}$ is again the total mass density of all gravitating constituents of the universe and $p_\mathrm{tot}$ their associated pressures which have to be taken into account in the frame of General Relativity as an additional source of gravitation. The general applicability of Eq. (\ref{36}) can be derived from the expression

\setlength{\mathindent}{0pt}
\begin{equation}
E=\int\limits_{}^\mathrm{3}(\rho_\mathrm{tot}c^\mathrm{2} +3p_\mathrm{tot})\sqrt{-g_\mathrm{3}}\mathrm{d}^\mathrm{3}V, \label{37}%
\end{equation}
where $\sqrt{-g_\mathrm{3}}\mathrm{d}^\mathrm{3}V$ is the local differential proper volume of space given through the determinant of the 3-D part of the Robertson-Walker metric tensor in the form

\setlength{\mathindent}{0pt}
\begin{equation}
\sqrt {-g_\mathrm{3}}  = \sqrt{-g_\mathrm{11}g_\mathrm{22}g_\mathrm{33}} = \frac{{R_\mathrm{u}^\mathrm{3}}}{{(1 - \frac{{kr^\mathrm{2}}}{4})^\mathrm{3}}}, \label{38}
\end{equation}
with k being the curvature parameter and r the normalized radial coordinate of a polar coordinate system. Then Eq. (\ref{37}) yields

\setlength{\mathindent}{0pt}
\begin{equation}
E = 4\pi R_\mathrm{u}^\mathrm{3} (\rho _\mathrm{tot} c^\mathrm{2}  + p_\mathrm{tot} )\int\limits_\mathrm{0}^\mathrm{1} {\frac{{r^\mathrm{2} dr}}
{{(1 - \frac{{kr^\mathrm{2} }}{4})^\mathrm{3} }}}. \label{39}%
\end{equation}
For a flat universe with $k=0$ the value of the integral is $1/3$ and Eq. (\ref{39}) simply yields an energy $E={\frac{{4\pi}}{3}}R_\mathrm{u}^\mathrm{3}(\rho_\mathrm{tot} c^\mathrm{2}+3p_\mathrm{tot})$ as claimed in Eq. (\ref{36}). The quantity

\setlength{\mathindent}{0pt}
\begin{equation}
U=-{\frac{{8\pi^\mathrm{2}G}}{{15}}}(\rho+{\frac{{3p}}{{c^\mathrm{2}}}})^\mathrm{2}R_\mathrm{u}^\mathrm{5} \label{40}%
\end{equation}
represents the negative potential gravitational binding energy which was
calculated by Fahr (2004) with the help of the well-known Poisson equation
for the cosmic gravitational potential $\Phi$ written for a homogeneous
universe. This potential in the neighborhood of an arbitrary space point is
obtained from the equation

\setlength{\mathindent}{0pt}
\begin{equation}
\Delta\Phi={\frac{1}{{r^\mathrm{2}}}}{\frac{\partial}{{\partial r}}}(r^\mathrm{2}%
{\frac{{\partial\Phi}}{{\partial r}}})=-4\pi G(\rho+{\frac{{3p}}{{c^\mathrm{2}}}}). \label{41}%
\end{equation}
In the above equations $c$ is the light velocity, $G$ is Newton's
gravitational constant, $\rho$ is the density of all contributing masses in
the universe (i.e. baryonic , dark and vacuum), and $p$ is the cosmic pressure
connected with these mass representer{s. The pressure contributions depend on
the associated types of mass density} (i.e. density of baryonic, dark or
vacuum-related equivalent masses).

In the following we assume that according to standard modern views the total
density $\rho$ is composed of $\rho_\mathrm{mat}$ due to baryonic and dark matter,
and of $\rho_\mathrm{vac}$ due to the mass equivalent of the energy density of the
cosmic vacuum, thus yielding the expression $\rho=\rho_\mathrm{mat}+\rho_\mathrm{vac}$.

To further evaluate the Eqs. (\ref{35}) through (\ref{40}), we have to
consider the cosmic pressures and their relation to densities. Here we base
ourselves on the well known thermodynamic relation used in cosmology and given by

\setlength{\mathindent}{0pt}
\begin{equation}
{\frac{\mathrm{d}(\rho c^\mathrm{2} R_\mathrm{u}^\mathrm{3} )}{{\mathrm{d}t}}} = - {\frac{\mathrm{d} (\varepsilon R_\mathrm{u}^\mathrm{3} )}{{\mathrm{d}t}}}
- p{\frac{\mathrm{d} (R_\mathrm{u}^\mathrm{3} )}{{\mathrm{d}t}}}. \label{42}%
\end{equation}
The above equation describes the change of the energy density with the
expansion of the cosmic scale $R_\mathrm{u}$ (Term 1) which is connected with the
corresponding change of the inner energy density $\epsilon$ (Term 2) and of
the work that is done by the cosmic pressure $p$ at the expansion of the
cosmic volume (Term 3), where one may remind that $p=p_\mathrm{mat}+p_\mathrm{vac}$. In
the present epoch of the cosmic evolution it is justified to assume that due
to adiabatic temperature decrease the inner energy density $\epsilon$ can be
neglected with respect to the rest mass energy density leading Eq. (\ref{42}) into the simpler form

{\setlength{\mathindent}{0pt}
\begin{equation}
{\frac{\mathrm{d}(\rho c^\mathrm{2} R_\mathrm{u}^\mathrm{3} )}{{\mathrm{d}R_\mathrm{u}}}} = - p{\frac{{\mathrm{d}R_\mathrm{u}^\mathrm{3}}}{{\mathrm{d}R_\mathrm{u}}}}. \label{43}%
\end{equation}
In the above relation we have converted derivatives with respect to $t$ into
ones with respect to $R_\mathrm{u}$ by setting $\mathrm{d}/\mathrm{d}t={(\mathrm{d}R_\mathrm{u}/\mathrm{d}t)\mathrm{d}/\mathrm{d}R_\mathrm{u}}$.

Representing all cosmic densities which in a homogeneous universe can only
depend on $R_\mathrm{u}$, by a general form of an $R_\mathrm{u}$- dependence according to $R_\mathrm{u}^\mathrm{{-n}}$,
one then obtains for the pressure $p$ from Eq. (\ref{43})

{\setlength{\mathindent}{0pt}
\begin{equation}
p=-{\frac{3{-n}}{3}}\rho c^\mathrm{2}. \label{44}%
\end{equation}
This result contains the well known relations of the thermodynamic pressure
$p$ and the density $\rho$ within the expanding universe and reveals the fact
that both pressure and density are described by the same dependence on the
cosmic scale $R_\mathrm{u}$. This gives the relations well known from general literature,
namely for the case of pure matter universe with: $p_\mathrm{mat}=0$ for $\rho
_\mathrm{mat}\sim R_\mathrm{u}^\mathrm{-3}$, for the case of pure photon pressure with: $p_\mathrm{\gamma
}=1/3\rho_\mathrm{\gamma}c^\mathrm{2}$ for $\rho_\mathrm{\gamma}\sim R_\mathrm{u}^\mathrm{-4}$, und for the case of a
pure cosmic vacuum with $p_\mathrm{vac}=-\rho_\mathrm{vac}c^\mathrm{2}$ for $\rho_\mathrm{vac}=const.$,
which can easily be checked by insertion of the respective pressures into
Eq. (\ref{43}).

While a pressure $p_\mathrm{mat}=0$ for adiabatically cooling matter, and a pressure
$p_\mathrm{\gamma}=1/3\rho_\mathrm{\gamma}c^\mathrm{2}$ for cosmologically redshifted photons seem
to be in accordance with our physical intuition, the negative pressure
$p_\mathrm{vac}=-\rho_\mathrm{vac}c^\mathrm{2}$ for the vacuum, however, appears to be
counterintuitive at first glance. One should, however, not forget that such a
result arises only in line with the assumption that according to (\ref{43}) the
equivalent mass density of the cosmic vacuum is constant at the expansion of the
universe. If vacuum energy density is constant then the expanding universe,
creating new cosmic volume, though doing work against the vacuum pressure also
permanently has to create new amounts of vacuum energy connected with the
increased volume. This in fact is only thermodynamically permissible, if the
vacuum pressure is negative according to Eq. (\ref{43}).

Based on these knowledges on the behaviour of cosmic density $\rho$ in an
expanding universe and of the associated cosmic pressure $p$ we now come back
to the Eqs. (\ref{35}) through (\ref{40}). To fulfill the requirement
$L=E+U=L_\mathrm{0}$ for the economical universe or, equivalent $\mathrm{d}L_0/\mathrm{d}R_\mathrm{u}=0$,

{\setlength{\mathindent}{0pt}
\begin{equation}
\frac{\mathrm{d}}{\mathrm{d}R_\mathrm{u}}[{\frac{{4\pi}}{3}}R_\mathrm{u}^\mathrm{3}c^{2}(\rho+{\frac{{3p}%
}{{c^\mathrm{2}}}})-{\frac{{8\pi^\mathrm{2}G}}{{15}}}(\rho+{\frac{{3p}}{{c^\mathrm{2}}}})^\mathrm{2}%
R_\mathrm{u}^\mathrm{5}]=0, \label{45}%
\end{equation}
exactly two solutions are existing. First, for the case $(\rho+{\frac
{3p}{c^\mathrm{2}}})\neq0$, one can obtain from Eq. (\ref{45}) the following condition

\setlength{\mathindent}{0pt}
\begin{equation}
(\rho+{\frac{{3p}}{{c^\mathrm{2}}}})={\frac{{5c^\mathrm{2}}}{{2\pi GR_\mathrm{u}^\mathrm{2}}}}. \label{46}%
\end{equation}
Second, for the case $(\rho+{\frac{3p}{c^\mathrm{2}}})=0$, one obtains%

\setlength{\mathindent}{0pt}
\begin{equation}
{p=-{\frac{1}{3}}\rho c^\mathrm{2}}. \label{47}%
\end{equation}
Interestingly enough, the requirements of Eqs. (\ref{46}) and (\ref{47})
are fulfilled simultaneously, as can easily be confirmed. Eq. (\ref{46})
states, that the density $\rho$ has to scale according to $R_\mathrm{u}^\mathrm{-2}$, while
Eq. (\ref{47}) yields a pressure $p=-{\frac{1}{3}}\rho c^\mathrm{2}$ , from
where one derives with Eq. (\ref{44}), that on the basis of the formal law
describing the $R_\mathrm{u}$-dependence according to $R_\mathrm{u}^\mathrm{-n}$, the necessary exponent has
to be $n=2.$ This, however, also implies that density as well as pressure
should scale with $R_\mathrm{u}^\mathrm{-2}$. From this behaviour of the mass
density with $R_\mathrm{u}$ one must, however, conclude that in a curvature-less universe
the total mass of the universe, i.e. $M={\frac{4\pi}{3}}\rho R_\mathrm{u}^\mathrm{3}$, does
linearly scale with the scale $R_\mathrm{u}$ yielding%

\setlength{\mathindent}{0pt}
\begin{equation}
M\sim R_\mathrm{u}. \label{48}%
\end{equation}
With the above result for an $L_\mathrm{0}=0$ - universe one obtains a nonclassical
connection of the world mass $M$ and the size, or scale, $R_\mathrm{u}$, of the universe.
This quasi-Machian form of a scaling of masses with the dimension of the
universe has, in fact already often been claimed for (e.g. Mach 1983;
Barbour 1995; Hoyle 1990, 1992; Thirring 1918; Barbour \& Pfister 1995).
The cosmological consequences of this surprising scaling law for the world
mass $M$ have been investigated in this paper. The $M\sim R_\mathrm{u}$ relation and the $\rho\sim R_\mathrm{u}^\mathrm{-2}$ relation surprisingly enough
have just most recently again been found from new formulations of logically
and physically rigorous concepts for what should be called an instantaneous
mass and diameter of an observer-related universe (Fahr \& Heyl 2006).

\section{Summary}
The assumed $R_\mathrm{u}^\mathrm{-2}$ scaling of matter and
vacuum energy densities combined with the assumption of a flat universe with curvature $k=0$ leads to a universe which does not face a horizon problem any longer
and thus does not require a cosmic inflation at the beginning. Furthermore,
the theoretically calculated and unexplainable high amount of vacuum energy -
with a value about $10^\mathrm{120}$ higher than observed - one perfectly fits into
the idea of an universe with a $R_\mathrm{u}^\mathrm{-2}$ scaling of the matter and vacuum energy density. In addition, it has been shown,
that the present universe might have its origin in a quantum mechanical
fluctuation that took place in the Planck era and that the vacuum energy is
nothing else but the scaling rest energy associated with half a Planck mass within a
Planck volume $V_\mathrm{Pl}$. Finally, the whole mass of the universe can be
explained by the accumulation of half Planck masses up to the present time which
are generated as virtual quantum mass releases per Planck time and permitted
to become real in the expanding universe. These results have been obtained on the basis of a so-called "economical" universe where the rest energy of the released Planck masses are compensated by the negative gravitational binding energy and which expresses one of the most fundamental laws of physics - the conservation of energy.

\end{document}